\documentclass{nature}

\usepackage{epsfig}
\newcommand{\gtrsim}{\lower.7ex\hbox{$\;\stackrel{\textstyle>}{\sim}\;$}}
\newcommand{\lesssim}{\lower.7ex\hbox{$\;\stackrel{\textstyle<}{\sim}\;$}}

\title{An infrared ring around the magnetar SGR$\,$1900$+$14}

\author{S. Wachter$^1$, E. Ramirez-Ruiz$^2$, V.\ V. Dwarkadas$^3$, 
C. Kouveliotou$^4$, J. Granot$^5$, S.\ K. Patel$^6$ \& D. Figer$^7$}

\begin{document}

\maketitle

\begin{affiliations}
\item Spitzer Science Center, California Institute of Technology, 
Pasadena, CA 91125, USA
\item Astronomy \& Astrophysics, 201 Interdisciplinary Sciences Building, 
Santa Cruz, CA 95064, USA
\item Department of Astronomy and Astrophysics, University of Chicago, 
5640 South Ellis Avenue, AAC 010c, Chicago, IL 60637, USA
\item NASA/Marshall Space Flight Center, 320 Sparkman Drive, 
Huntsville, AL 35805, USA
\item Centre for Astrophysics Research, University of Hertfordshire, 
College Lane, Hatfield, Herts, AL10 9AB, UK
\item Optical Sciences Corporation, 6767 Old Madison Pike, Suite 650, 
Huntsville, AL 35806, USA
\item Chester F. Carlson Center for Imaging Science, Rochester Institute of Technology, 
54 Lomb Memorial Drive, Rochester, NY 14623, USA
\end{affiliations}

\begin{abstract}
Magnetars\cite{duncan92, woods04} are a special class of
slowly rotating (period $\sim$5-12 s) neutron stars with extremely strong
magnetic fields ($>$10$^{{\bf 14}}$ G) -- at least an order of magnitude
larger than those of the ``normal'' radio pulsars.  The potential
evolutionary links and differences between these two types of objects
are still unknown; recent studies, however, have provided
circumstantial evidence connecting magnetars with very massive progenitor stars
\cite{figer05, gaensler05, muno06}. Here we report the
discovery of an infrared elliptical ring or shell surrounding the
magnetar SGR$\,$1900$+$14.  The appearance and energetics of the ring
are difficult to interpret within the framework of the
progenitor's stellar mass loss or the subsequent evolution of the supernova remnant.
We suggest instead that a dust-free cavity was produced in the
magnetar environment by the giant flare emitted by the source in
August 1998. Considering the total energy released in the flare, the theoretical
dust--destruction radius matches well with the observed dimensions of the ring.  
We conclude that SGR$\,$1900$+$14 is unambiguously associated
with a cluster of massive stars, thereby solidifying
the link between magnetars and massive stars.


\end{abstract}

Soft gamma repeaters (SGRs) and anomalous X-ray pulsars (AXPs) are the
two main classes of objects currently believed to be magnetars.  Their
characteristic ages, derived from their rotational
properties\cite{chryssa98}, indicate a young population -- typically a
few thousand years old. Although these spin-down ages must be treated
with caution, the relative youth of SGRs and AXPs is supported by the
fact that some AXPs are located at the centres of supernova remnants
(SNRs). Associations between SNRs and SGRs have also been claimed in the
literature, but, unlike the AXPs,
the SGRs are offset from the centres of their proposed SNRs, increasing
the likelihood of spurious alignment. As a result, the validity of all
SGR--SNR associations has been questioned\cite{gaensler01}.
Both AXPs and SGRs, for example the AXP CXOU J164710.2$-$455216 (ref.5)
and SGR$\,$1806$-$20 (ref.8), 
have been linked to clusters of
massive stars. The implied progenitor mass of $M \geq 40-50 M_{\odot}$
supports theoretical predictions that very massive stars with
sufficient metallicity and the corresponding high mass loss can
still form neutron stars\cite{heger03}, contrary to the standard
evolutionary picture that such massive stars end their lives as black
holes.

As part of our systematic study of the circumstellar environments of
SGRs and AXPs, we observed the position of SGR$\,$1900$+$14 using all three
instruments onboard the NASA Spitzer Space Telescope in 2005 and 2007.
Surprisingly, our 16$\,\mu$m and 24$\,\mu$m images (Fig. 1b, c)
reveal a prominent ring-like structure that is not detected in our
3.6--8.0$\,\mu$m observations.  A formal elliptical fit to the ring
indicates semi-major and semi-minor axes of angular lenghts $\sim\,$36'' and
$\sim\,$19'', respectively, centred at right ascension 19:07:14.32 and
declination
+09:19:20.0 $\pm$1.0''.  This coincides with the radio position of
SGR$\,$1900$+$14 (19:07:14.33 +09:19:20.1 $\pm$0.15''), determined
from observations of a transient synchrotron source associated with
the 1998 giant flare\cite{frail99}.  We re-examined published and
archival data of the field around SGR$\,$1900$+$14 but found no
equivalent structure at optical, near-IR, radio or X-ray
wavelengths. In particular, the lack of detection in the radio spectrum, to a limit
of~\cite{kaplan02} $L_{\rm 332 MHz} \leq 4 \times
10^{29}\;d_{15}^2\;{\rm erg\;s^{-1}}$, and the X-ray spectrum, to a limit of $L_{\rm 2-10 keV}
\leq 2.7 \times 10^{33}\;d_{15}^2\;{\rm erg\;s^{-1}}$ (Chandra data
archive, http:\/\/cxc.harvard.edu\/cda\/), is important in
determining the nature of the ring. Here $d_{15}$ parameterizes the distance to the source, 
according to $d = 15\;d_{15}\;$kpc.

The Spitzer images are dominated by the bright emission from two
nearby M5 supergiants that mark the centre of a compact cluster of
massive stars at a distance of 12--15$\;$kpc\cite{vrba96,vrba00}.
Although it has been suggested that SGR$\,$1900$+$14 is associated with
this cluster\cite{vrba96}, an alternative distance of 5 kpc has also been suggested,
on the basis of the hydrogen column density of its X-ray
spectrum\cite{hurley99}. In addition, the SGR is offset by 12'' from the
cluster centre, and the visual extinction corresponding to the X-ray
derived hydrogen column density\cite{kaplan02} ($A_V=12.8\pm0.8$) is
significantly different from that deduced optically for the cluster
stars\cite{vrba96} ($A_V=19.2\pm 1$). Considering both possibilities,
we calculated physical sizes for the semi-major and semi-minor axes of
the ring of $0.9\times 0.5\;$pc, if SGR$\,$1900$+$14 lies at a distance
of $\sim 5\;$kpc, or $2.6\times 1.4\;$pc, if it resides at a cluster
distance of $15\;$kpc.

We have constrained the temperature of the material in the ring by
first cross-convolving the images with the point spread function of
the different arrays. We then measured the flux through fixed apertures at
several positions along the ring. The largest source of uncertainty
lies in the subtraction of the local background. We derived a
temperature of $130-150\;$K for the material in the ring using a
simple black--body fit, and one of $\sim 80-120\;$K using a more
realistic dust model.\cite{draine} We caution that the presence of
spectral lines or peculiarities in the underlying mid-infrared spectral
shape might significantly alter these derived temperatures.  For
example, Spitzer observations have revealed a shell only visible at
24$\,\mu$m, where the broadband flux can be entirely attributed to
[OIV] line emission\cite{morris06}.  A broad 22$\,\mu$m continuum
feature has also been identified in environments associated with
supernovae and massive star formation\cite{chan00}.

The 70$\,\mu$m image shows a bipolar flux distribution along the minor
axis of the ellipse (Fig. 1d). Extended enhancements in the ring are
also seen at these positions in the 16 and 24$\,\mu$m images.  Our data
do not allow us to determine the underlying three--dimensional geometry of the
structure, that is, to distinguish between a limb-brightened shell, a true
ring morphology, or a bipolar cavity with equatorial torus.
Similarly, an accurate measurement of the total flux in the ring is very
difficult because the enveloping diffuse emission (Fig. 1a) coupled with 
the contribution from the M5 supergiants 
prohibits a clean separation of the ring
emission. Using a narrow elliptical aperture, we measured fluxes of
1.2$\pm$0.2$\;$Jy and 0.4$\pm$0.1$\;$Jy at 24$\,\mu$m and 16$\,\mu$m,
respectively, implying ring luminosities of $\nu L_\nu(16\mu m)\sim 2 \times
10^{36}\;d_{15}^2\;{\rm erg\;s^{-1}}$ and $\nu L_\nu(24\mu m)\sim 4 \times 10^{36}\;d_{15}^2\;{\rm
  erg\;s^{-1}}$ (where $\nu$ denotes the frequency at which the luminosity is determined).

To test for ring evolution, we used the publicly
available MIPSGAL Spitzer Legacy programme observations of the field
that were obtained at an epoch (2005 Oct 03) earlier than was our own data
set. Creating a difference image from the 24$\,\mu$m data of the two
different epochs, we find no discernable change in the size
(positional shifts $< 0.5$'') or flux (root mean square $< 1-2$\%) of the ring
emission in the 1.6 yr between the two observations.  The
stationary nature of the ring strongly disfavours one obvious explanation
for the creation of the ring -- a light echo, due to reprocessing by
the surrounding dust, from the 1998 SGR$\,$1900$+$14 giant flare.  This
model predicts a typical fractional change in the ring size and
brightness between the two epoch observations of $\sim 23\%$, which is
not seen in the data (see Supplementary Information for more details).

Morphologically, the ring resembles the wind-blown bubbles and shells
observed around evolved massive stars\cite{smith07, weis03, gruendl00}, such as
B supergiants, luminous blue variables and Wolf-Rayet stars. The
supersonic wind from the star drives a shock into the surrounding
medium, sweeping up the ambient material into a thin, dense, cool
shell. The emission from the swept-up material is powered by the
luminous central star. Such shells have typical radii of $\sim
2-10\;$pc for Wolf-Rayet stars\cite{gruendl00} and $0.1-2.3\;$pc for luminous blue variables/supergiants\cite{weis03, smith07}
, overlapping the size of our ring. To guard against a simple
chance superposition with an unrelated source, we
investigated the photometric properties of stars
near the centre of the ring. The sources in the 
immediate vicinity of the SGR position\cite{kaplan02} are too faint 
in comparison with the near-infrared luminosities expected for 
the massive stars discussed above\cite{crowther07,cox00},
and thus cannot be physically associated with the ring. For stars with 
larger offsets from the centre ($\geq 7$'') we cannot rule out a massive star
classification based solely on available archival data.
However, a situation in which an off-centre star creates an elliptical ring 
with an unrelated magnetar at its exact centre 
appears rather contrived.  

In the absence of an obvious chance superposition, the close positional
coincidence between the SGR and the centre of the ring then strongly
indicates that the SGR and the ring are physically connected. One 
interpretation is that the ring represents material ejected during the
late stages of the SGR progenitor's evolution. However, we know that a
supernova explosion occurred, producing the magnetar, and the
resulting shock wave is expected to disrupt such a close shell as it
interacts with and dissipates into the surrounding material. 
Some cases are known in which the SNR has over-run a wind-swept shell, 
but the wind bubble
invariably emerges irregular and fragmented in shape after its
encounter with the supernova\cite{kooheiles95}. The symmetric and well-defined
appearance of our ring thus would imply that 
the supernova shock wave is still inside the ring. Our
simulations show that this would require the supernova to be $< 200$
yr old, with bright radio and X-ray
emission, contrary to observations. The lack of high-energy X-ray and
non-thermal radio emission also excludes the possibility that the
observed ring is produced by the emission of the blast wave from the
progenitor's supernova explosion.

The rejection of these various possibilities raises the issue of the
energy source that is powering the ring.  We derived a limit for the
required luminosity of the heating source by means of radiation
balance, using the measured temperature $T_{\rm IR}$ of the dust. 
If the luminosity
$L_\ast$ is dominated by a source of temperature $T_\ast$, then
(see Supplementary Information for details)
\begin{equation}
L_\ast \approx 2.7 \times 10^{40} T_{\rm IR}^5 T_\ast^{-1} R^2\;{\rm erg\;s^{-1}}
\end{equation}
where $T_{\rm IR}$ is measured in units of 100 K, $T_\ast$ is measured in
units of $10^{4}$ K and $R$, the distance from the central heating source, is 
measured in units of parsecs. 

We note that the persistent X-ray luminosity of the SGR is only $L_{\rm
2-10\,{\rm keV}} \approx 2.0-3.5 \times 10^{35}\;d_{15}^2\;{\rm
erg\;s^{-1}} \ll L_\ast$ and, thus, cannot generate the observed infrared 
emission from the ring.  The only viable heat source consistent with the
observed properties of the ring appears to be irradiation from the
stars in the nearby cluster. Figure 1a clearly shows that both the
cluster and SGR$\,$1900$+$14 are embedded in a diffuse, extended cloud
of 24$\,\mu$m emission.  Similar diffuse 24$\,\mu$m emission is also
observed in association with the clusters containing
SGR$\,$1806$-$20 (ref.24) and Westerlund 1, implying that the
extended emission is powered by the cluster stars. We note that, given 
the location and size of the 24$\,\mu$m nebula, the cluster might be 
more spatially extended than previously thought, with 
as-yet-undiscovered member stars.

A possible mechanism for the creation of the ring is the formation of a dust-free cavity
owing to the giant flare activity from SGR$\,$1900+14, which would naturally explain why the 
ring is centred on the magnetar. 
The destruction of dust grains can occur either by sublimation due to the heating by 
ultraviolet\cite{WD00} or X-ray\cite{FKR01} emission, or by grain charging due to the incident X-rays, 
which causes the dust grains to gradually shatter into smaller pieces until they are eventually destroyed\cite{FKR01}. 
The size and shape of the ring
can constrain the energy output and degree of
anisotropy of the flaring event 
(see Supplementary Information for details). Using Equation (25) of ref.26 with the
fiducial values and $\alpha =0$ (where the flux at frequency $\nu$ is proportional to $\nu^{-\alpha}$), we find that 
the dust destruction radius matches well with the observed dimensions of
the ring around SGR$\,$1900$+$14 for $E \geq 6\times
10^{45}d_{15}^2\;$erg. This estimate is close to the observed
isotropic equivalent energy\cite{tanaka07} in the initial spike of the
27 August 1998 flare; the latter, however, is only accurate for photon energies $\geq
50\;$keV, whereas the estimated dust-destruction radius requires photons with
energies of $\sim 1$ keV. 

Alternatively, a previous, more
energetic giant flare from SGR$\,$1900$+$14, for example, similar to the
  2004 December 27 giant flare from
  SGR$\,$1806-20 (refs 27, 28), with $E \approx
10^{47}d_{15}^2\;$erg, could have carved the dust-free cavity during
the magnetar's estimated spin-down age of $\sim 1,800\;$yr. For a rate of one
flare per $\sim 50$ yr (ref.28), it is reasonable to expect
at least one such event $10^3t_{\rm kyr}\;$yr ago,
with $t_{\rm kyr} \approx1$. Then, for the current location of the SGR still to
be at the centre of the ring to a precision of 1'',
the proper motion of the magnetar (due
to its birth kick velocity $v_\perp$) and the proper motion of the
dust-free cavity (with velocity $v_c$) should satisfy
$\max(v_c,v_\perp) \leq
71d_{15}t_{\rm kyr}^{-1}\;{\rm km\;s^{-1}}$.

In light of the probable association between the SGR and the star
cluster, we have re-examined the issue of the mismatch in extinction,
which has been used as an argument against that
connection\cite{kaplan02}.  As our observations show, the cluster-SGR
environment is characterized by a complex dust distribution on small
spatial scales and dust destruction by the high-energy emission of the
SGR. Because the derived extinction values were based on
measurements towards two distinct physical locations (the supergiants
and the SGR) and are based on different tracers (gas and dust)
it is then not surprising that the resulting values differ. Similar
differences have been observed for the Cas A SNR\cite{hartmann97}
and extragalactic gamma ray burst sources\cite{stratta05}. In
addition, we note that the visual absorption derived solely from the near-infrared
photometry of the M supergiants, using both the published
data\cite{vrba96} and new measurements extracted from the Two Micorn All Sky Survey, is
A$_V= 10.4 - 14.4$ mag, in comparison with  A$_V=19.2\pm 1$ mag, the value obtained  when the optical I band
measurement is included. Although this is puzzling and deserves
further investigation, the various extinction measurements are clearly
subject to large uncertainties and can potentially be reconciled with
our conclusion that the SGR is a member of the star cluster.










\begin{addendum}
 \item This work is based on observations made with the Spitzer Space
   Telescope, which is operated by the Jet Propulsion Laboratory,
   California Institute of Technology under a contract with NASA.
   Support for this work was provided by NASA through an award issued
   by JPL/Caltech.  This publication also makes use of data products
   from the Two Micron All Sky Survey, which is a joint project of the
   University of Massachusetts and the Infrared Processing and
   Analysis Center/California Institute of Technology, funded by the
   National Aeronautics and Space Administration and the National
   Science Foundation.  J.G. gratefully acknowledges a Royal Society
   Wolfson Research Merit Award. VVD acknowledges support from the
   National Science Foundation, and useful discussions with R. McCray, A. 
   Crotts and R. Chevalier. D.F. acknowledges support from NASA through the 
   Long Term Space Astrophysics programme, and by the New York  State Foundation for Science,
   Technology, and Innovation Faculty Development Program grant.  
 \item[Competing interests statement] The authors declared no competing interests. 
 \item[Supplementary information] accompanies this paper. 
 \item[Correspondence] Correspondence and requests for materials
should be addressed to S.W. (wachter@ipac.caltech.edu).
\end{addendum}


\newpage
\begin{center}
{\bf \Large Supplementary Material for ``An infrared ring around the
  magnetar SGR$\,$1900$+$14''}
\end{center}

\section*{Light echo scenario details}\label{flare}
If the SGR were surrounded by an extensive dust cloud, a substantial
amount of the giant flare radiation will be converted into IR
radiation. Given the extended nature of the dust cloud, light travel
times across it must be considered. The two MIPS 24$\;\mu$m images
were obtained at $t_1 = 7.10\;$yr and $t_2 = 8.75\;$yr after the 1998
August 27 giant flare from SGR$\,$1900+14. If the ring is from a light
echo due to reprocessing of the giant flare photons then the time
delay $t$ is related to the location of the (assumed rapid)
reprocessing by $ct = r(1-\cos\theta)$, where $(r,\theta,\phi)$ are
polar coordinates with the SGR located at the origin and with the
$z$-axis pointing to the observer. This locus of points from which
reprocessed photons reach the observer simultaneously describes a
paraboloidal surface centered on the location of the SGR and symmetric
about our line of sight to the SGR, which grows linearly in time, in
the sense that $r/t$ is a function of the polar angle $\theta$
alone. The lateral distance from the SGR to this surface (at $\theta =
90^\circ$) is simply $ct$, which corresponds to $ct_1 \approx 2.2\;$pc
and $ct_2 \approx 2.7\;$pc for our two epochs of observation, where
$ct_1$ corresponds to the semi-major axis of the ring for $d \approx
12.5\;$kpc. 

The main characteristics of such a light echo can be demonstrated by
the simple and idealized case of a thin spherical shell of dust of
radius $R$ centered on the SGR. Its intersection with the paraboloidal
surface described above is simply at $\cos\theta = 1-ct/R$ so that the
apparent ring size $R_{\rm ring}$ is
\begin{equation}
R_{\rm ring} = R\sin\theta = R\sqrt{\frac{ct}{R}\left(2-\frac{ct}{R}\right)}
\quad\quad 0 < \frac{ct}{R} < 2\ .
\end{equation}
It initially grows super-luminally, its expansion slowing down until it
reaches a maximal size of $R_{\rm ring} = R$ at $t = R/c$, and then
begins to shrink increasingly fast until it disappears at $t = 2R/c$.
Its apparent velocity $\beta_{\rm app}$ in units of $c$ is
\begin{equation}
\beta_{\rm app} = \frac{1}{c}\frac{dR_{\rm ring}}{dt} =
\frac{1-ct/R}{\sqrt{\frac{ct}{R}\left(2-\frac{ct}{R}\right)}} \quad\quad 0 <
\frac{ct}{R} < 2\ .
\end{equation}
While $\beta_{\rm app} \geq 1$ for much of the time, we also have
$\beta_{\rm app} \ll 1$ at $t \approx R/c$, i.e. very close to the
time when the ring reaches its maximal size and begins to
shrink. However, for two generic observation times one expects a
fractional change of $\sim(t_2-t_1)/t_1$ in the ring size, which in
our case corresponds to $\sim 23\%$. Moreover, since the observed ring
in our case is elongated with an axis ratio of $\sim 2:1$, the shell
of reprocessing dust must clearly be non-spherical, and therefore the
maximal size would also depend on the azimuthal angle $\phi$, so that
the apparent expansion velocity of the ring would never vanish
everywhere at once, and one can expect fractional changes of the order
of $\sim(t_2-t_1)/t_1 \sim 23\%$ in the ring shape or brightness
between the two epochs, at least in some parts of the ring. Therefore,
our strict limits on its exact location, and fractional chances in its
size, shape, and brightness (of $<1-2\%$) are very hard to satisfy
simultaneously in this scenario.

\section*{Dust heating considerations}

The temperature of the dust can be used to determine the required
luminosity of the heating source by means of radiation balance. If
$\epsilon_\lambda$ is the dust absorption efficiency, then its
temperature is given by a balance between the heat input due to the
flux it absorbs and emits. If the flux it absorbs is dominated by a
point source (which is also a useful approximation for a group of
sources at distances larger than their separations) with luminosity
$L_{\ast}$ peaking at $\lambda_{\ast}$, at a distance $R$, then a
(spherical, for simplicity) dust grain of radius $a$ absorbs energy at
a rate of
\begin{equation}
\dot{E}_{\rm absorbed} = \epsilon_{\lambda_{\ast}}\pi a^2 \frac{L_\ast}{4\pi R^2}\ ,
\end{equation}
and radiates away energy at a rate of
\begin{equation}
\dot{E}_{\rm radiated} = 4\pi a^2 \epsilon_{\lambda_{\rm IR}} \sigma T_{\rm IR}^4\ ,
\end{equation}
where $\lambda_{\rm IR} \sim h c / k T_{\rm IR}$. Equating the two
rates gives:
\begin{equation}\label{lum0}
L_\ast =16 \pi R^2 \left(\frac{\epsilon_{\lambda_{\rm IR}}}
{\epsilon_{\lambda_{\ast}}}\right)\sigma T_{\rm IR}^4\ .
\end{equation}
If the luminosity $L_\ast$ is dominated by a source of temperature
$T_\ast$ ($\sim hc/k\lambda_{\ast}$), and both
$\epsilon_{\lambda_{\ast}}$ and $\epsilon_{\lambda_{\rm IR}}$ are in
the range where $\epsilon_\lambda \propto 1/\lambda$, then
\begin{equation}
\frac{\epsilon_{\lambda_{\rm IR}}}{\epsilon_{\lambda_{\ast}}} \sim
\frac{T_{\rm IR}}{T_\ast} = 0.01 \left(\frac{T_{\rm IR}}{100\;{\rm
K}}\right) \left(\frac{T_{\ast}}{10^{4}\;{\rm K}}\right)^{-1}\ .
\label{t}
\end{equation}
and
\begin{equation}
L_\ast = 2.7 \times 10^{40} \left(\frac{T_{\rm IR}}{100\;{\rm K}}\right)^5
\left(\frac{T_\ast}{10^4\;{\rm K}} \right)^{-1}
\left(\frac{R}{1\;{\rm pc}}\right)^2\;{\rm erg\;s^{-1}}\ .
\label{lum}
\end{equation}

This basically sets a limit for the required heating source. Note that
for a given source equation (\ref{lum}) implies that the dust
temperature decreases rather slowly with the distance from the source,
$T_{\rm IR} \propto R^{-2/5}$ and could help explain why we do not
detect a change along the ring in the apparent temperature of the ring emission.

\section*{Giant flare energetics and anisotropy}
The $\sim 2:1$ axis ratio of the ring suggests that if it was
indeed formed by the initial spike of a giant flare, then the flare had an
angular variation in its isotropic equivalent luminosity $L$ or energy
$E$ of a factor of $\geq 4$ (since the sublimation radius scales as
$L^{1/2}$ while the dust shattering radius scales as $E^{1/2}$, and
the true axis ratio of the dust-free cavity could be somewhat larger
than the observed axis ratio of the ring due to projection
effects). Moreover, the relatively modest degree of anisotropy
suggests that the true total energy output in the event that created
the dust-free cavity was $\geq 6 \times 10^{45}d_{15}^2\;$erg, which
is comparable to the {\it isotropic equivalent} energy output for the
2004 giant flare from SGR$\,$1806-20. For the latter event it is hard
to estimate the true energy output, since we have no observations that
directly constrain the emission in directions other than our line of
sight.

\newpage
\begin{figure}
\centerline{
            \epsfig{file=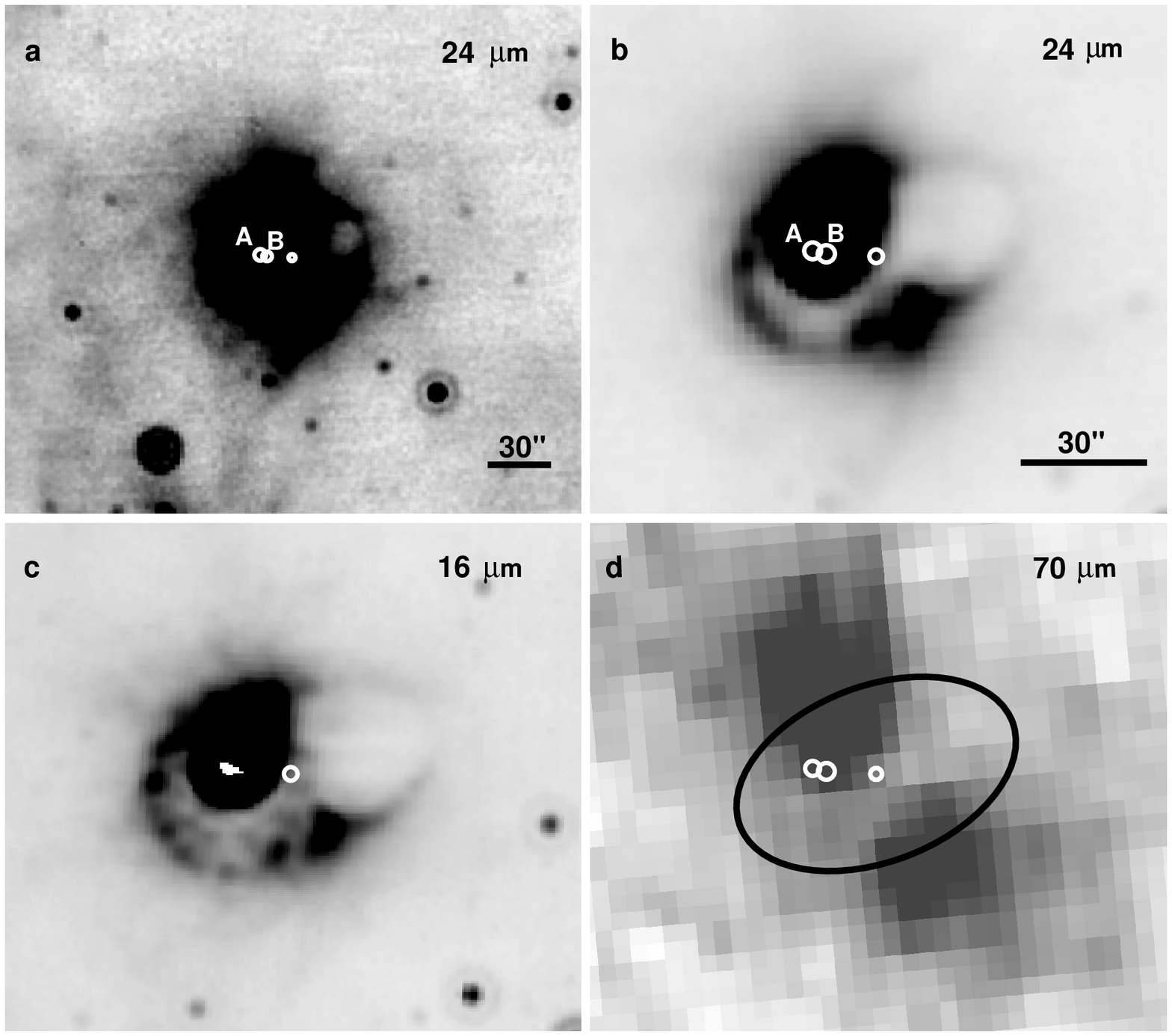, scale=0.9}
            }
\caption{{\bf The infrared ring around SGR$\,$1900$+$14.}  Spitzer
Space Telescope imaging of SGR$\,$1900$+$14 was acquired on 2005
September 20 using the InfraRed Array Camera (3.6$-$8.0$\,\mu$m),
on 2005 Oct 11 with the blue (16$\,\mu$m) peak-up imaging mode of the
InfraRed Spectrograph (IRS), and on 2007 May 29 with the Multiband
Imaging Photometer for Spitzer (MIPS; $24, 70\mu$m).  Only the MIPS
and IRS images are shown here. 
We assembled the basic calibrated data products
into combined images using the Mosaicking and Point-source Extraction package. For the MIPS
data, we first applied a differential flat field to correct for small
(1$-$2\%) instrumental artefacts. The final combined images 
({\bf a--d}) have effective exposure times of 31 s for IRS and 16 s
for both MIPS channels. The white circle in the centre of each image marks the radio
position of the SGR. For all images, north is up and east is to the left.
{\bf a}, MIPS 24$\,\mu$m image scaled
to show the diffuse emission enveloping both the SGR and the nearby
cluster. The white circles labeled `A' and `B' indicate the positions of
the red supergiants\cite{vrba96} located at the centre of the star cluster.  
{\bf b}, Close-up of the MIPS
24$\,\mu$m image shown in {\bf a}, but scaled to highlight the
ring-like structure of the extended emission.  {\bf c}, IRS
16$\,\mu$m peak-up image covering the same field of view as {\bf b}.
{\bf d}, MIPS 70$\,\mu$m image showing emission associated
with the enhanced areas along the minor axis of the ellipse also seen
at 16 and 24$\mu$m. The position of the ring is indicated
by the black ellipsoid.  The convolution of the 16 and 24$\,\mu$m images to the
resolution of the 70$\,\mu$m array indicates that the ring emission is
too spatially confined and faint to be detectable at this wavelength.
}
\end{figure}

\end{document}